\documentclass{webofc}
\usepackage[varg]{txfonts}   
\woctitle{Physics at the Magnetospheric Boundary}
\begin{document}
\title{Emission line diagnostics for accretion and outflows in young very low-mass stars and brown dwarfs}
%
%

\author{B.Stelzer\inst{1}\fnsep\thanks{\email{stelzer@astropa.inaf.it}} \and
        J.M.Alcal\'a\inst{2} 
        \and  E.Whelan\inst{3} 
        \and  A.Scholz\inst{4} 
}

\institute{INAF - Osservatorio Astronomico di Palermo, Piazza del Parlamento 1, 90134 Palermo, Italy
\and
           INAF - Osservatorio Astronomico di Capodimonte, Via Moiariello 16, 80131 Napoli, Italy
\and
           Institut f\"ur Astronomie \& Astrophysik, Eberhard-Karls-Universit\"at T\"ubingen, 72076 T\"ubingen, Germany
\and
           School of Physics \& Astronomy, University of St. Andrews, The North Haugh, St. Andrews, Fife\,KY16\,9SS, United Kingdom
          }

\abstract{%
We discuss accretion and outflow properties 
of three very low-mass young stellar objects based on broad-band 
mid-resolution X-Shooter/VLT spectra. 
Our targets (FU\,Tau\,A, 2M\,1207-39, and Par-Lup3-4) 
have spectral types between M5 and M8, ages between $1$\,Myr and $\sim 10$\,Myr, 
and are known to be accreting from previous studies.  
The final objective of our project is the determination of mass
outflow to accretion rate for objects near or within 
the substellar regime as a probe for the T\,Tauri phase of brown dwarfs
and the investigation of variability in the accretion and outflow processes. 
}
\maketitle
\section{Introduction}\label{intro}

According to the paradigm of accretion in low-mass young stellar objects (YSOs) the 
inner-disk material is channeled along magnetic field lines onto the star 
\citep{Koenigl91.1} where it 
dissipates its kinetic energy in standing shocks \citep{Calvet98.0}. 
Jets and molecular outflows are believed to be intimately related to the accretion process,
carrying away the angular momentum. 
The ratio between mass outflow and mass accretion rate 
($\dot{M}_{\rm out}/\dot{M}_{\rm acc}$) is, therefore, a key parameter of jet
launching models \citep{Pudritz05.0, Ferreira06.0}. For T\,Tauri stars values in a range
of $\approx 0.01 - 0.1$ have been measured \citep{Hartigan95.1, Melnikov09.0}. Only
few estimates have been obtained for the faint end of the stellar sequence where preliminary
results suggest the $\dot{M}_{\rm out}/\dot{M}_{\rm acc}$ ratio to be significantly 
higher (e.g. \cite{Whelan09.0}). The sensitive latest generation spectrographs enable
to extend studies of the accretion/outflow connection into the brown dwarf regime.


We present results of broad-band mid-resolution ($350-2500$\,nm, $R \sim 9000$)
X-Shooter/VLT spectroscopy for three YSOs. 
The data were obtained in the framework of the Italian GTO {\it Survey
of nearby galactic star forming regions} \citep{Alcala11.0}. 
X-Shooter spectroscopy 
allows for a detailed characterization of YSOs, 
including an accurate assessment of their fundamental parameters, kinematics, 
rotation, and magnetic activity. 
It provides a rich database of accretion diagnostics from the 
Br$\gamma$ and Pa$\beta$ lines in the near-IR to the Balmer jump in the UV 
including the full optical band with the Balmer series and He\,$\lambda$5876 
and the Ca\,IRT. Finally, outflows can be traced through forbidden line 
emission. 

In this contribution we focus on three of the most interesting individual objects 
from our survey. The characteristics and a motivation for their selection
as X-Shooter targets are given in Sect.~\ref{sect:sample}. We describe the methods to
derive mass accretion and outflow rates in Sect.~\ref{sect:methods}. The results are
presented in Sect.~\ref{sect:results}, conclusions and open questions are summarized in 
Sect.~\ref{sect:outlook}.

\section{Targets}\label{sect:sample}

More than $80$ YSOs were observed with X-Shooter in our survey of 
low-mass galactic star forming regions. 
Here we describe the results for three objects with particularly interesting or
puzzling properties in the previous literature. 
The target list is presented in Table~\ref{tab1}. 

\begin{table}
\centering
\caption{Target properties and X-shooter observing log. Additional data for Par-Lup3-4
is described in Sect.~\ref{subsect:results_parlup34}.}
\label{tab1}       
\begin{tabular}{lccccccc}\hline
Name       & Region & distance & SpT & $M_*$ & $R_*$ & Obs.date & Exp. \\
           &        & [pc]     &     & [$M_\odot$] & [$R_\odot$] & & [s] \\ \hline
FU\,Tau\,A  & Taurus & 140 & M6.5 & 0.08 & 1.4 & Jan 11, 2011 & $1800$ \\
2M\,1207-39 & TWA & 53 & M8 & 0.035 & 0.21 & Mar 22, 2010 / Apr 19, 2012 & $3600$ \\
Par-Lup3-4  & Lupus\,III & 200 & M5 & 0.13 & 0.18 & Apr 7, 2010 & $3600$ \\
\hline 
\end{tabular}
\end{table}

{\bf FU\,Tau\,A} is the primary in a wide brown dwarf binary system 
with $5.7^{\prime\prime}$ separation corresponding to $\approx 800$\,AU at
the distance of Taurus \citep{Luhman09.1}. Various signatures of youth
include the presence of a circumstellar disk for both components inferred
from near-IR excess emission and ongoing accretion inferred from H$\alpha$ 
emission. The binary is located in the 
Barnard\,215 dark cloud with no other young star nearby. Its isolation 
makes FU\,Tau a benchmark object for brown dwarf formation scenarios. 
FU\,Tau\,A appears well above the youngest
isochrone of evolutionary pre-main sequence models in the HR diagram. 
FU\,Tau\,B, also appears younger than $1$\,Myr but by a smaller amount \citep{Luhman10.0}. 
One objective for observing FU\,Tau\,A with X-Shooter was a better characterization
of its accretion given the evidence for X-ray emission produced 
by an accretion shock from a {\em Chandra} observation \citep{Stelzer10.0}.

{\bf 2M1207-39} is the most well-studied among the brown dwarfs in the
nearby $\approx$\,10\,Myr old TW\,Hya association. 
It has a disk from which it is accreting \citep{Scholz05.4}, 
an outflow \citep{Whelan12.0} and a planetary companion
\citep{Chauvin04.1}. 
An almost two orders of magnitude variation of the accretion rate has been inferred 
in the past based on the H$\alpha$ emission from different epochs \citep{Stelzer07.2}. 
This strong variability and the perception of 2M\,1207-39 as a kind of prototype accreting
brown dwarf have led us to include it in our X-Shooter survey.

{\bf Par-Lup3-4} is a very low-mass star in the Lupus\,III cloud discovered by \cite{Comeron03.1}. 
It is underluminous by $\approx\,4$\,mag with respect to other young M5 stars in the same
star forming complex. This can be explained by an edge-on disk. In fact, a disk inclination
angle of $\sim 81^\circ$ was inferred from SED modelling \citep{Huelamo10.0}. 
The jet of Par-Lup3-4 was imaged in optical forbidden lines \citep{Fernandez05.1}. 
Our multi-epoch X-Shooter spectroscopy aims at studying the 
accretion and outflow variability and the relation of the mass accretion and
mass outflow rates. The high inclination of the disk is believed to be favorable for
the detection of variations related to non-azimuthal structures in the 
accretion geometry.

\section{Measurement of mass accretion and outflow rates}\label{sect:methods}

We use three methods to measure mass accretion rates, $\dot{M}_{\rm acc}$,  
from the X-Shooter spectra. 
First, we use the empirical relation with the full width of the H$\alpha$ line at 
$10$\,\% of the peak height, 
$\log{\dot{M}_{\rm acc}}\,{\rm [M_\odot/yr]} = (-12.89 \pm 0.3) + (9.7 \pm 0.7) \cdot 10^{-3} \cdot W_{10\%}\,{\rm [km/s]}$; 
see \cite{Natta04.2}. 
Secondly, we apply empirical correlations between the luminosity of individual emission
lines ($L_{\rm line}$) and the accretion luminosity of the type 
$L_{\rm acc} = a + b \cdot L_{\rm line}$ where $a, b$ are coefficients that have been
presented e.g. by \cite{Herczeg08.1}.
These correlations have been newly calibrated for an unprecedentedly large number of lines
on the basis of the X-Shooter spectra of low-mass stars in Lupus obtained within our
GTO program \citep{Alcala13.0}. Finally, we measure the excess emission with respect
to a purely photospheric spectrum. This is done by adding to the spectrum of a 
non-accreting template star, that has the same spectral type as the target, the emission 
of an 
isothermal slab model for a hot hydrogen gas. The model parameters (temperature, density
and length of the slab) are adapted until the template plus model give the best fit to
the target spectrum. This provides the
accretion luminosity, $L_{\rm acc}$, which can be transformed into mass accretion rate
according to $\dot{M}_{\rm acc} = (1 - 1/R_{\rm in})^{-1} \cdot \frac{L_{\rm acc} R_* }{G M_*}$
where $R_*$ and $M_*$ are the stellar radius and mass, $G$ the gravitational constant, and
$R_{\rm in}$ the inner disk radius, assumed here to be $5\,R_*$  
(see \cite{Gullbring98.1}).  

Outflow rates are measured from line luminosities using the prescription given by
\cite{Hartigan95.1}. In particular, we use the relation 
$\log{\dot{M}_{\rm out}} = X \cdot ( 1 + \frac{n_{\rm c}}{n_{\rm e}} ) (\frac{V_{\rm T}}{150\,{\rm km/s}}) {(\frac{l_{\rm T}}{2 \cdot 10^{15}\,{\rm cm}})}^{-1} (\frac{L_{\rm line}}{L_\odot})$.
Here, $n_{\rm c}$ is the ion-specific critical density and $n_{\rm e}$ the electron density.
The projected aperture size on the plane of the sky ($l_T$) is determined by the slit 
width and the distance to the target. The tangential velocity ($V_T$) can be obtained 
from the measured radial velocity if the inclination angle of the outflow is known. 
$X$ is a numerical constant calculated by \cite{Hartigan95.1} for the most prominent 
forbidden lines. 

\begin{figure}[t]
\centering
\includegraphics[width=14.5cm,clip]{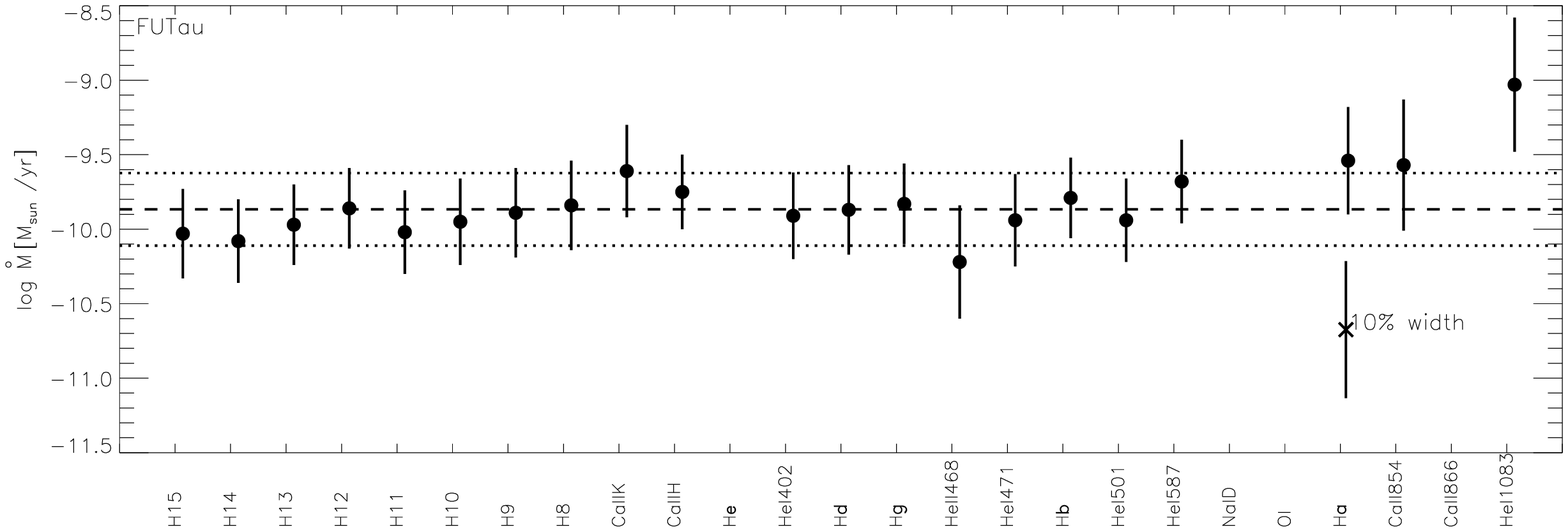}
\caption{Mass accretion rate for FU\,Tau\,A derived from various emission line fluxes
and luminosities using new calibrations from \protect\cite{Alcala13.0}, 
and from the H$\alpha$ $10$\,\% width. Dashed and dotted lines are mean and
standard deviation for the luminosity-based $\dot{M}_{\rm acc}$ excluding 
lines with suspected wind contribution, blending or uncertain calibration.}
\label{fig1}       
\end{figure}

\section{Results}\label{sect:results}

\subsection{FU\,Tau\,A}\label{subsect:results_futau}

In Fig.~\ref{fig1} we show the values for $\dot{M}_{\rm acc}$ that we obtain from the
individual emission line luminosities using the calibrations of \cite{Alcala13.0}
and from the H$\alpha$ $10$\,\% width applying the relation from \cite{Natta04.2}. 
While the various 
line luminosities yield consistent values within the uncertainties, the width of H$\alpha$
results in a lower estimate for $\dot{M}_{\rm acc}$. 
From the slab model we obtain a third value for $\dot{M}_{\rm acc}$. That 
latter estimate is 
$\log{\dot{M}_{\rm acc}}\,{\rm [M_\odot/yr]} = -10.1$ and it agrees well with the result 
from the line luminosities. The discrepancy of the value obtained from the H$\alpha$ width
supports previous notions of the poor reliability of this tracer as quantitative
measure for accretion (e.g. \cite{Herczeg08.1,Costigan12.0,Alcala13.0}).

We have found for the first time outflow activity in FU\,Tau\,A, detected in the form of 
forbidden emission lines, and we use the [OI]$\lambda630.0$\,nm and 
[SII]$\lambda673.1$\,nm emission to derive the mass loss rate, $\dot{M}_{\rm out}$. 
The values of $\dot{M}_{\rm out}$ ($\approx 4 \cdot 10^{-11}\,{\rm M_\odot/yr}$),
and consequently $\dot{M}_{\rm out}/\dot{M}_{\rm acc} \approx 0.3$,  
are not well constrained because of uncertainties in the electron density and the 
inclination angle required to calculate the tangential velocity.
Specifically, as a result of the non-detection of [S\,II]$\lambda 617.7$\,nm emission
only a lower limit can be determined for $n_{\rm e}$. Furthermore, the disk inclination 
angle required to calculate $V_T$ is also a lower limit, $i_{\rm disk} \geq 50^\circ$ 
from a comparison of the 
measured $v \sin{i}$ and the rotation period observed photometrically by 
\cite{Scholz12.2}; see also \cite{Stelzer13.0} for details.  
The outflow of FU\,Tau\,A has recently also been detected at millimeter wavelengths 
\citep{Monin13.0}, and this estimate of the molecular outflow rate is an order of 
magnitude higher than that obtained from the atomic transitions.

\begin{figure*}[t]
\centering
\includegraphics[width=14.5cm,clip]{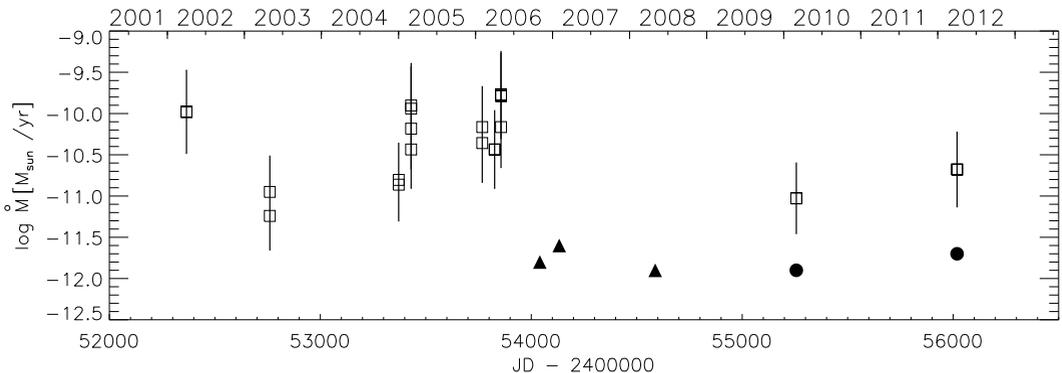}
\caption{Mass accretion rate of 2M\,1207-39 measured throughout the last decade with different
methods: line luminosity (filled circles), modelling of UV excess (filled triangles) and
H$\alpha$ $10$\,\% width (open squares). For the latter one the plotting symbols denote
the highest and the lowest values measured in the respective month and the vertical bars
include the uncertainties of the conversion to mass accretion rate. 
The two most recent epochs shown in this graph are from our X-Shooter observations.}
\label{fig:mdot_jd_2m1207}       
\end{figure*}

\subsection{2M\,1207-39}\label{subsect:results_2m1207}

2M\,1207-39 is one of the best-studied young accreting brown dwarfs. We have collected 
accretion measurements from the literature. The resulting time-series for 
$\dot{M}_{\rm acc}$ is shown in Fig.~\ref{fig:mdot_jd_2m1207}. Different plotting symbols
refer to different techniques for calculating the accretion rate (see figure caption). 
Our X-Shooter observations are represented by the two most recent epochs. 
It can be noticed that all
values derived from the H$\alpha$ width yield systematically higher $\dot{M}_{\rm acc}$
than other methods (UV excess modelling and line luminosity). 
Our simultaneous measurement of H$\alpha$ width and line luminosities 
(the two most recent epochs in Fig.~\ref{fig:mdot_jd_2m1207}) prove 
that there are discrepancies associated with the different analysis methods which,
if ignored, could be mistaken for true accretion variations. As noted above for the
case of FU\,Tau\,A UV excess and line luminosity are considered more reliable quantitative
accretion diagnostics. All data points obtained so far for 2M\,1207-39 from these two tracers 
suggest at most modest variability of the accretion rate on time-scales of years.

\subsection{Par-Lup3-4}\label{subsect:results_parlup34}

\begin{figure}
\centering
\parbox{14.5cm}{
\parbox{8.5cm}{
\includegraphics[width=8.5cm,clip]{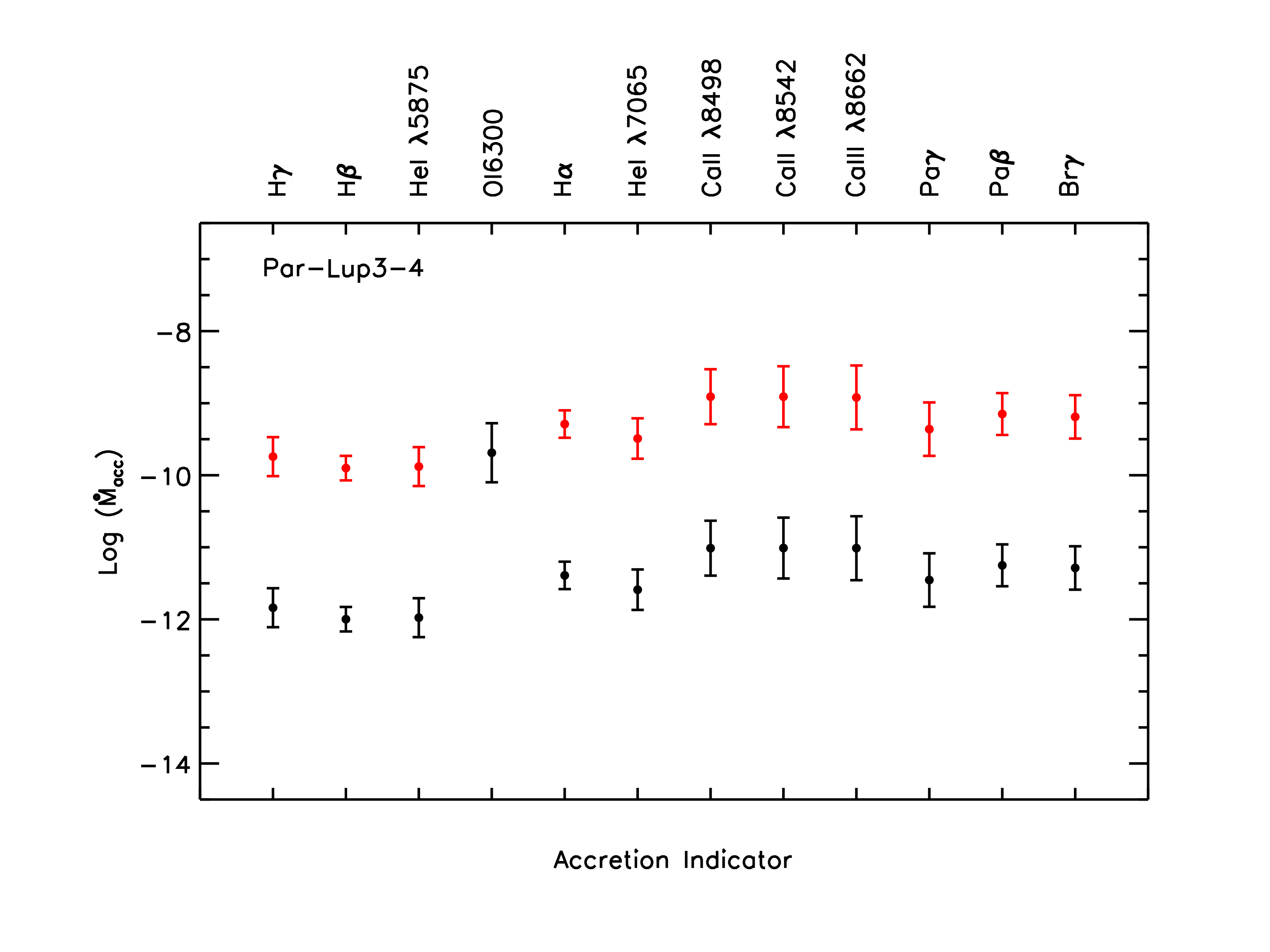}
}
\parbox{6cm}{
\includegraphics[width=6cm,clip]{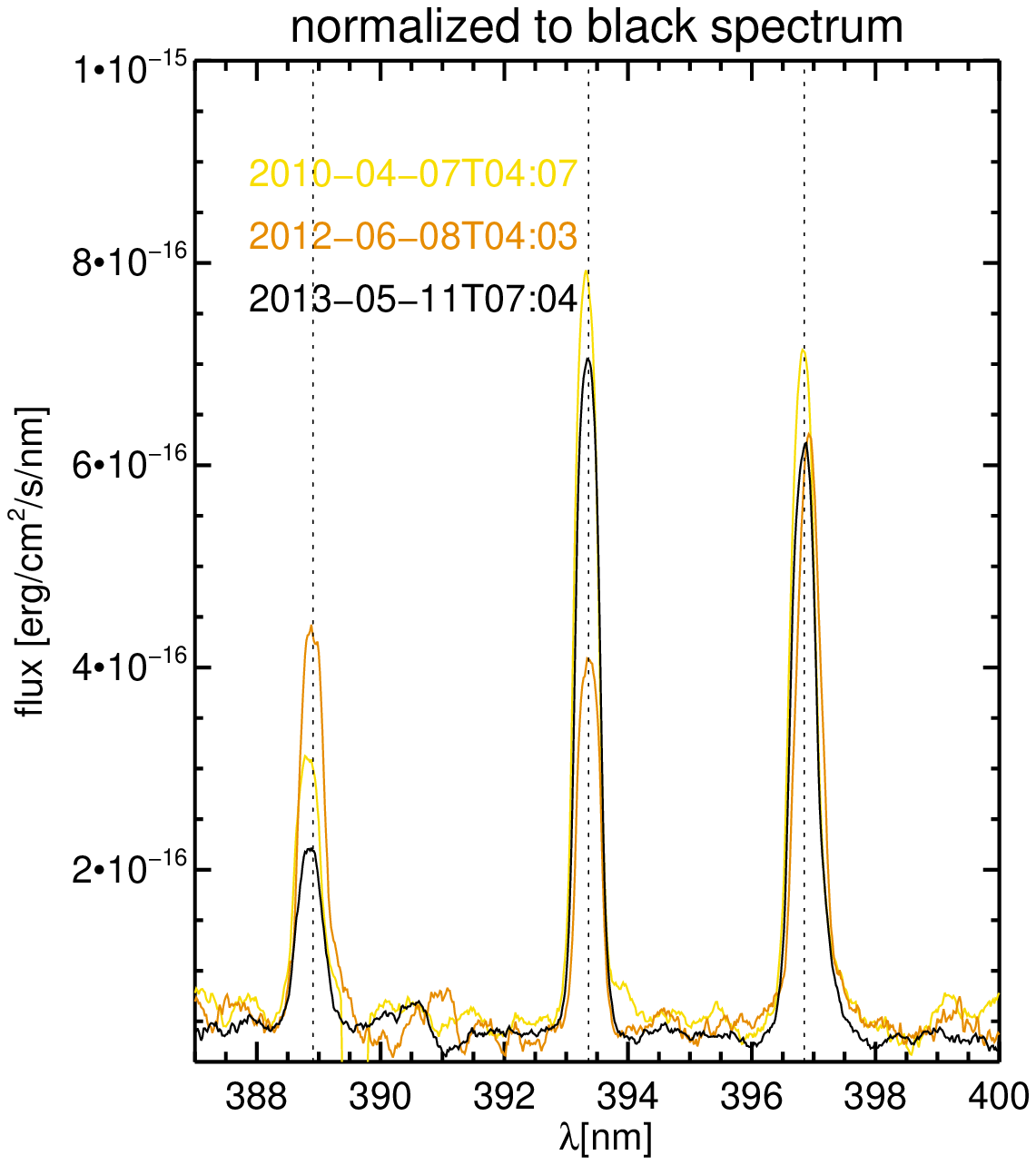}
}
}
\caption{Results from X-Shooter spectroscopy of Par-Lup3-4: (Left) - Mass accretion rates
calculated from line luminosities with the calibrations of \cite{Alcala13.0} (black symbols)
and same measurements corrected for the grey extinction induced by the edge-on disk (red
symbols); see \protect\cite{Whelan13.0} for details. (Right) -  Three different epochs 
showing apparently uncorrelated changes in the emission line fluxes. Shown are the spectrum 
obtained in the GTO program (yellow) and two spectra from our follow-up programs carried
out in June 2012 and May 2013.}
\label{fig:parlup34}       
\end{figure}

The mass accretion rates of Par-Lup3-4 derived from the line
luminosities with the calibrations from \cite{Alcala13.0} are shown in the left panel of 
Fig.~\ref{fig:parlup34}. The actual measurements (shown in black) need to be corrected
for the grey extinction induced by the edge-on disk of Par-Lup3-4. Assuming that the same
obscuration factor ($OF$) suppresses both the bolometric luminosity and the accretion luminosity, $\dot{M}_{\rm acc}$
can be corrected according to $(\dot{M}_{\rm acc})_{\rm corr} = OF^{1.5} \cdot \dot{M}_{\rm acc}$.
This is based on the relation between accretion rate and luminosity given in 
Sect.~\ref{sect:methods} and considers the $R_*^2$ proportionality of the bolometric 
luminosity; see \cite{Whelan13.0} for more details. The corrected values for 
$\dot{M}_{\rm acc}$ are shown in red in Fig.~\ref{fig:parlup34}. 
No obscuration correction is applied to the [OI]$\lambda630$\,nm emission because 
forbidden lines are generated in outflows not in accretion streams. The good agreement
between the $\dot{M}_{\rm acc}$ derived from [OI]$\lambda630$\,nm with the extinction corrected
values obtained for the other lines confirms (1) that the outflow is not obscured by the
disk and (2) that there is a tight connection between outflow and accretion activity.  
%

In June 2012 and May 2013 we obtained in total nine additional X-Shooter spectra of 
Par-Lup3-4 with the aim of studying its emission line variability. Thus, the available
spectra for this object cover timescales from few hours to three years. 
Our preliminary analysis 
shows evidence for changes in the flux distribution from years 2010 to 2013, 
e.g. the June 2012 spectrum displays the strongest Balmer H8 line but the weakest
Ca\,II\,H\&K lines in Fig.~\ref{fig:parlup34}.
The changes in emission line flux are accompanied by
variations in the spectral slope (not shown). 
A detailed investigation of all available spectra will
shed light on variations on different timescales.

\section{Conclusions and open questions}\label{sect:outlook}

For all our observations, but particularly demonstrated for 2M\,1207-39, 
the H$\alpha$ $10$\,\% width yields mass accretion rates that 
are inconsistent with
results from other methods, possibly due to the influence of outflows on the line profiles.
The case of Par-Lup-3-4 shows the importance of an accurate assessment of extinction for
measuring $\dot{M}_{\rm acc}$ and, consequently, the mass 
inflow-to-outflow ratio. For the example of FU\,Tau\,A we discuss the difficulties in 
obtaining accurate $\dot{M}_{\rm out}$ estimates from forbidden lines. Future sensitive 
observations of larger samples of very low-mass objects will also have to address the
discrepancies between outflow rates determined from atomic and molecular transitions.
%
%
\bibliography{xshooterbds}
%

\end{document}